\documentclass[3p,times,10pt]{elsarticle}
\usepackage{amsfonts}
\usepackage{amssymb}
\usepackage{amsmath}
\usepackage{graphicx}
\usepackage{epsfig}
\usepackage{ctable}
\usepackage{float}

\begin{document}

\begin{frontmatter}

\title{Parallel dynamics between non-Hermitian and Hermitian systems}
\author{P.~Wang, S.~Lin, L.~Jin and Z.~Song\corref{cor1}}
\ead{songtc@nankai.edu.cn}
\cortext[cor1]{Corresponding author}
\address{School of Physics, Nankai University, Tianjin 300071, China}

\begin{abstract}
We study the connection between a family of non-Hermitian Hamiltonians $%
\mathcal{H}$ and Hermitian ones $H$\ based on exact solutions. In general,
for a dynamic process in a non-Hermitian system $\mathcal{H}$, there always
exists a parallel dynamic process governed by the corresponding Hermitian
conjugate Hamiltonian $\mathcal{H}^{\dag }$. We show that a linear
superposition of the two parallel dynamics is exactly equivalent to the time
evolution of a state under a Hermitian Hamiltonian $H$. It reveals a novel
connection between non-Hermitian and Hermitian systems.
\end{abstract}


\end{frontmatter}

\section{Introduction}

When speaking of the physical significance of a non-Hermitian Hamiltonian,
it is implicitly assumed that there exists another Hermitian Hamiltonian
which shares the complete or partial spectrum with the non-Hermitian
Hamiltonian \cite{Bender 98,Bender 99,Dorey 01,Dorey 02,Bender
02,A.M43,A.M36,Jones,M.Z40,M.Z41,M.Z82}. Mostafazadeh proposed a
metric-operator method to compose a Hermitian Hamiltonian, which has exactly
the same real spectrum with the pseudo-Hermitian Hamiltonian \cite{A.M}.
From the Hermitian counterpart, one can extract the physical meaning of a
pseudo-Hermitian Hamiltonian in\ the viewpoint of spectrum \cite%
{A.M38,A.M391,A.M392,JLPT}. Alternatively, in previous works \cite%
{L.J81,L.J83,L.J44}, we established a connection between a non-Hermitian
Hamiltonian and an infinite Hermitian system\ in the viewpoint of eigen
state. However, this connection does not provide the link between the
dynamics of the two systems.

In this work, we study the connection between a $\mathcal{PT}$ non-Hermitian
Hamiltonian and a Hermitian one by linking the dynamics in the systems. We
consider a group of Hamiltonians $\{H,\mathcal{H},\mathcal{H}^{\dag }\}$ on
the same lattice, where $H$ is Hermitian and $\mathcal{H}$\ is
non-Hermitian. It is shown that $H$ and $\mathcal{H}$\ may share\ a common
subset of eigenvalues, while the corresponding eigenfunction of $H$\ can be
written as the superposition of the ones from $\mathcal{H}$\ and $\mathcal{H}%
^{\dag }$. Since the connection is a type of set to set, it allows the
equivalence between the dynamics of the two systems. We note that, for a
dynamic process in a non-Hermitian system, there always exists a parallel
dynamic process\ governed by the corresponding Hermitian conjugate
Hamiltonian. It is shown that a linear superposition of the two parallel
dynamics may be exactly equivalent to the time evolution of a state under a
Hermitian Hamiltonian. It reveals a novel connection between non-Hermitian
and Hermitian systems.

This paper is organized as follows. In Sec.\ \ref{General formalism}, we
present the general formalism. Section \ref{Illustrative examples} is
devoted to demonstrate the main idea by illustrative examples. In Sec.\ \ref%
{Parallel dynamics}, we apply the obtained result on the dynamics of states
in relevant systems. Finally, we give a summary and discussion in Sec.\ \ref%
{Conclusion and discussion}.

\section{General formalism}

\label{General formalism}

\begin{figure}[tbp]
\centering
\includegraphics[ bb=33 279 481 686, width=0.4\textwidth, clip]{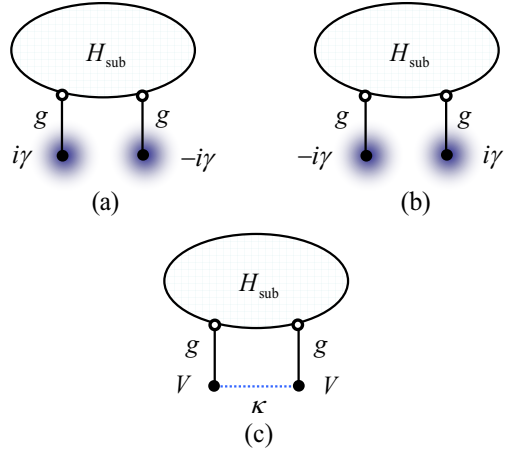}
\par
\caption{ Schematic illustration of configurations for non-Hermitian models
and their Hermitian correspondence. (a) is the graph for a non-Hermitian
Hamiltonian $\mathcal{H}$, where $H_{\text{\textrm{sub}}}$ is an arbitrary
Hermitian sub-system. (b) is the graph for the non-Hermitian Hamiltonian $%
\mathcal{H}^{\dag }$. (c) is the corresponding Hermitian graph, which may
have a connection with $\mathcal{H}$ and $\mathcal{H}^{\dag }$.}
\label{fig1}
\end{figure}

We start our investigation by considering a class of Hermitian and
non-Hermitian Hamiltonians $\{H,\mathcal{H},\mathcal{H}^{\dag }\}$, which
consist of two parts
\begin{eqnarray}
\mathcal{H} &=&H_{\text{\textrm{sub}}}+H_{\gamma }, \\
H &=&H_{\text{\textrm{sub}}}+H_{\kappa }.
\end{eqnarray}%
The structure of the systems is schematically illustrated in Fig. \ref{fig1}%
. Here $H_{\text{\textrm{sub}}}$ describes a Hermitian tight-binding
Hamiltonian on an arbitrary graph, and non-Hermitian term $H_{\gamma }$ and
Hermitian term $H_{\kappa }$ are in the form%
\begin{eqnarray}
H_{\gamma } &=&g(\left\vert a\right\rangle \left\langle A\right\vert
+\left\vert b\right\rangle \left\langle B\right\vert +\mathrm{H.c.})  \notag
\\
&&-i\gamma \left\vert A\right\rangle \left\langle A\right\vert +i\gamma
\left\vert B\right\rangle \left\langle B\right\vert
\end{eqnarray}%
and%
\begin{eqnarray}
&&H_{\kappa }=g(\left\vert a\right\rangle \left\langle A\right\vert
+\left\vert b\right\rangle \left\langle B\right\vert +\mathrm{H.c.})  \notag
\\
&&+\kappa (\left\vert A\right\rangle \left\langle B\right\vert +\mathrm{H.c.}%
)+V(\left\vert A\right\rangle \left\langle A\right\vert +\left\vert
B\right\rangle \left\langle B\right\vert ),
\end{eqnarray}%
where $\left\vert a\right\rangle $ $(\left\vert b\right\rangle )$ and $%
\left\vert A\right\rangle $ $(\left\vert B\right\rangle )$ are the position
states at sites $a$ $(b)$\ in $H_{\text{\textrm{sub}}}$\ and $A$ $(B)$,
respectively. In this paper, the hopping integral $\kappa $\ and on-site
potential $V$\ are real. The wave functions of Hamiltonians $\{H,\mathcal{H},%
\mathcal{H}^{\dag }\}$ are in the forms%
\begin{equation}
\left\vert \eta \right\rangle =\left\vert \eta _{\mathrm{sub}}\right\rangle
+\eta _{a}\left\vert a\right\rangle +\eta _{b}\left\vert b\right\rangle
+\eta _{A}\left\vert A\right\rangle +\eta _{B}\left\vert B\right\rangle ,
\end{equation}%
where $\eta =\psi $, $\varphi $, and $\phi $ denotes the wave functions in
the three systems.

The Schr\"{o}dinger equations with the same real eigenenergy $\varepsilon $
are $H\left\vert \psi \right\rangle =\varepsilon \left\vert \psi
\right\rangle $, $\mathcal{H}\left\vert \varphi \right\rangle =\varepsilon
\left\vert \varphi \right\rangle $, and $\mathcal{H}^{\dag }\left\vert \phi
\right\rangle =\varepsilon \left\vert \phi \right\rangle $, which have the
explicit forms
\begin{equation}
\left\{
\begin{array}{c}
H_{\text{\textrm{sub}}}\left\vert \psi _{\mathrm{sub}}\right\rangle
=\varepsilon \left\vert \psi _{\mathrm{sub}}\right\rangle \\
g\psi _{a}-\varepsilon \psi _{A}+V\psi _{A}+\kappa \psi _{B}=0 \\
g\psi _{b}-\varepsilon \psi _{B}+V\psi _{B}+\kappa \psi _{A}=0%
\end{array}%
\right. ,  \label{S1}
\end{equation}%
and%
\begin{equation}
\left\{
\begin{array}{c}
H_{\text{\textrm{sub}}}\left\vert \varphi _{\mathrm{sub}}\right\rangle
=\varepsilon \left\vert \varphi _{\mathrm{sub}}\right\rangle \\
g\varphi _{a}-\varepsilon \varphi _{A}-i\gamma \varphi _{A}=0 \\
g\varphi _{b}-\varepsilon \varphi _{B}+i\gamma \varphi _{B}=0%
\end{array}%
\right. ,  \label{S2}
\end{equation}%
and%
\begin{equation}
\left\{
\begin{array}{c}
H_{\text{\textrm{sub}}}\left\vert \phi _{\mathrm{sub}}\right\rangle
=\varepsilon \left\vert \phi _{\mathrm{sub}}\right\rangle \\
g\phi _{a}-\varepsilon \phi _{A}+i\gamma \phi _{A}=0 \\
g\phi _{b}-\varepsilon \phi _{B}-i\gamma \phi _{B}=0%
\end{array}%
\right. ,  \label{S3}
\end{equation}

respectively. Combining the above two equations, we have%
\begin{equation}
\left\{
\begin{array}{c}
H_{\text{\textrm{sub}}}(\left\vert \varphi _{\mathrm{sub}}\right\rangle
+\left\vert \phi _{\mathrm{sub}}\right\rangle )=\varepsilon (\left\vert
\varphi _{\mathrm{sub}}\right\rangle +\left\vert \phi _{\mathrm{sub}%
}\right\rangle ) \\
g(\varphi _{a}+\phi _{a})-\varepsilon (\varphi _{A}+\phi _{A})-i\gamma
(\varphi _{A}-\phi _{A})=0 \\
g(\varphi _{b}+\phi _{b})-\varepsilon (\varphi _{B}+\phi _{B})+i\gamma
(\varphi _{B}-\phi _{B})=0%
\end{array}%
\right. .  \label{S23}
\end{equation}%
Comparing the Eqs. (\ref{S1}) and (\ref{S23}), we find that one can have%
\begin{equation}
\left\vert \psi \right\rangle =\left\vert \varphi \right\rangle +\left\vert
\phi \right\rangle ,  \label{superposition}
\end{equation}%
if the following conditions are satisfied%
\begin{equation}
\left\{
\begin{array}{c}
V\psi _{A}+\kappa \psi _{B}=-i\gamma (\varphi _{A}-\phi _{A}) \\
V\psi _{B}+\kappa \psi _{A}=i\gamma (\varphi _{B}-\phi _{B})%
\end{array}%
\right. ,  \label{conditions}
\end{equation}%
From Eq. (\ref{superposition}), we have%
\begin{equation}
\psi _{A}=\varphi _{A}+\phi _{A},\text{ }\psi _{B}=\varphi _{B}+\phi _{B}.
\end{equation}%
Submitting the above equations into Eq. (\ref{conditions}), we have%
\begin{equation}
\left(
\begin{array}{c}
\varphi _{A} \\
\varphi _{B}%
\end{array}%
\right) =-\frac{V+\kappa \sigma _{x}-i\gamma \sigma _{z}}{V+\kappa \sigma
_{x}+i\gamma \sigma _{z}}\left(
\begin{array}{c}
\phi _{A} \\
\phi _{B}%
\end{array}%
\right) ,
\end{equation}%
where $\sigma _{x}$ and $\sigma _{x}$ are Pauli matrices. It indicates that
the solutions of $\varphi _{A,B}$ and $\phi _{A,B}$ for a given $\gamma $\
may lead to the restriction on parameters $V$\ and $\kappa $.

In this work, the non-Hermiticity arises from the imaginary potentials. Then
we have $\mathcal{H}^{\dag }=\mathcal{H}^{\ast }$, which allows us to write
the eigenfunctions in the form of $\varphi _{A}=\phi _{A}^{\ast }$ and $%
\varphi _{B}=\phi _{B}^{\ast }$ for real-energy eigenstates. Therefore the
above equation can be reduced as
\begin{equation}
(V+\kappa \sigma _{x})\left(
\begin{array}{c}
\text{Re}\varphi _{A} \\
\text{Re}\varphi _{B}%
\end{array}%
\right) =\gamma \sigma _{z}\left(
\begin{array}{c}
\text{Im}\varphi _{A} \\
\text{Im}\varphi _{B}%
\end{array}%
\right) ,
\end{equation}%
or the explicit form%
\begin{equation}
\left\{
\begin{array}{c}
V\text{Re}\varphi _{A}+\kappa \text{Re}\varphi _{B}-\gamma \text{Im}\varphi
_{A}=0 \\
\kappa \text{Re}\varphi _{A}+V\text{Re}\varphi _{B}+\gamma \text{Im}\varphi
_{B}=0%
\end{array}%
\right. ,  \label{Eq_AB}
\end{equation}%
based on which one can establish the relation among $V$, $\kappa$, and $%
\gamma $. We see that the relation depends on the eigenstates of $\mathcal{H}
$.

For a $\mathcal{PT}$-symmetric system, which corresponds to parity symmetric
$H_{\text{\textrm{sub}}}$,\ a real-energy state can always be written as the
form of $\varphi _{A}=\pm \varphi _{B}^{\ast }$, i.e., Re$\varphi _{A}=\pm $%
Re$\varphi _{B}$ and Im$\varphi _{A}=\mp $Im$\varphi _{B}$. This leads to%
\begin{equation}
\left( V\pm \kappa \right) \text{Re}\varphi _{A}-\gamma \text{Im}\varphi
_{A}=0,  \label{Eq_A}
\end{equation}%
which means that the Hermitian Hamiltonian is\ $\gamma $-dependent.
Furthermore, for a simpler case with real $\varphi _{A}$ and $\varphi _{B}$,
the fact Im$\varphi _{A,B}=0$ leads to
\begin{equation}
\kappa =\pm V,
\end{equation}

and%
\begin{equation}
\varphi _{A}=\phi _{A}=\psi _{A}/2=\mp \varphi _{B}=\mp \phi _{B}=\mp \psi
_{B}/2,
\end{equation}%
correspondingly. Therefore if one can find a non-Hermitian system $\mathcal{H%
}$, which has an eigenvector with real components $\varphi _{A}=\mp \varphi
_{B}$, there should exist a Hermitian system $H$ with $\kappa =\pm V$ and
there must have an eigenvector in the form of Eq. (\ref{superposition}). And
the two eigenvectors have the same real eigenvalue.

Finally, it is necessary to stress that the relation expressed in Eq. (\ref%
{superposition})\ is subtle. If one find that the superposition $\left\vert
\varphi \right\rangle +\left\vert \phi \right\rangle $ corresponds to an
eigenstate $\left\vert \psi \right\rangle =\left\vert \varphi \right\rangle
+\left\vert \phi \right\rangle $ of a Hamiltonian $H$, another superposition
$\alpha \left\vert \varphi \right\rangle +\beta \left\vert \phi
\right\rangle $ may correspond to an eigenstate $\left\vert \psi ^{\prime
}\right\rangle =\alpha \left\vert \varphi \right\rangle +\beta \left\vert
\phi \right\rangle $ of another Hamiltonian $H^{\prime }$. Nevertheless, all
four states $\left\vert \varphi \right\rangle ,$ $\left\vert \phi
\right\rangle ,$ $\left\vert \psi \right\rangle $, and $\left\vert \psi
^{\prime }\right\rangle $ have the same eigenvalue with respect to their own
Hamiltonians $\mathcal{H},$ $\mathcal{H}^{\dag },$ $H$, and $H^{\prime }$,
respectively.

\begin{figure}[tbp]
\centering
\includegraphics[ bb=16 263 493 686, width=0.4\textwidth, clip]{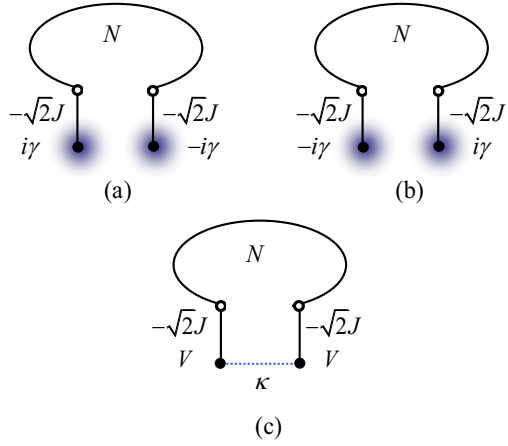}
\par
\caption{ Schematic illustrations for the uniform chain. Here $H_{\text{%
\textrm{sub}}}$ is a uniform chain of length $N$. Exact solution shows that
the combination of eigenstates of systems (a) and (b) is also the eigenstate
of system (c) under the condition $\protect\kappa =\pm V$.}
\label{fig2}
\end{figure}

\section{Illustrative examples}

\label{Illustrative examples}

\subsection{Uniform chain}

In this section, we investigate a simple and exactly solvable system to
illustrate the main idea of our paper. In order to exemplify the above
mentioned analysis of connecting the eigenstate of $H$\ to those of $%
\mathcal{H}$\ and $\mathcal{H}^{\dag }$, we take $H_{\text{sub}}$\ to be the
simplest network: a uniform chain. The sub-Hamiltonian in the sample
Hamiltonian has the form%
\begin{equation}
H_{\text{sub}}=-J\overset{N-1}{\sum_{l=1}}\left( \left\vert l\right\rangle
\left\langle l+1\right\vert +\mathrm{H.c.}\right) ,  \label{H_sub}
\end{equation}%
and non-Hermitian term%
\begin{eqnarray}
H_{\gamma } &=&-\sqrt{2}J(\left\vert 1\right\rangle \left\langle
A\right\vert +\left\vert N\right\rangle \left\langle B\right\vert +\mathrm{%
H.c.})  \notag \\
&&-i\gamma \left\vert A\right\rangle \left\langle A\right\vert +i\gamma
\left\vert B\right\rangle \left\langle B\right\vert ,  \label{H_gamma}
\end{eqnarray}%
and Hermitian term%
\begin{eqnarray}
&&H_{\kappa }=-\sqrt{2}J(\left\vert 1\right\rangle \left\langle A\right\vert
+\left\vert N\right\rangle \left\langle B\right\vert +\mathrm{H.c.})  \notag
\\
&&+\kappa (\left\vert A\right\rangle \left\langle B\right\vert +\mathrm{H.c.}%
)+V(\left\vert A\right\rangle \left\langle A\right\vert +\left\vert
B\right\rangle \left\langle B\right\vert ),  \label{H_kappa}
\end{eqnarray}%
which are sketched in Fig. \ref{fig2}. We note that $\mathcal{H}$($\mathcal{H%
}^{\dag }$)\ has $\mathcal{PT}$ symmetry, i.e., $\mathcal{PTH(\mathcal{H}%
^{\dag })}(\mathcal{PT})^{-1}=\mathcal{H}$($\mathcal{H}^{\dag }$), which was
proposed and exactly solved in Ref. \cite{L.J44}. Here $\mathcal{P}$ and $%
\mathcal{T}$\ represent the space-reflection operator (or parity operator)
and the time-reversal operator, respectively. The corresponding Hermitian
Hamiltonian $H$ has both $\mathcal{P}$ and $\mathcal{T}$ symmetries.

To demonstrate our result, we consider the solutions of the Hamiltonians
with $N=2$. On the basis $\{\left\vert A\right\rangle ,$ $\left\vert
l=1,2\right\rangle ,$ $\left\vert B\right\rangle \}$, the Hamiltonians can
be written as%
\begin{equation}
H=J\left(
\begin{array}{cccc}
V/J & -\sqrt{2} & 0 & \kappa /J \\
-\sqrt{2} & 0 & -1 & 0 \\
0 & -1 & 0 & -\sqrt{2} \\
\kappa /J & 0 & -\sqrt{2} & V/J%
\end{array}%
\right) ,
\end{equation}%
and%
\begin{eqnarray}
&&\mathcal{H}=\mathcal{H}^{\dag }(-\gamma )  \notag \\
&=&J\left(
\begin{array}{cccc}
-i\gamma /J & -\sqrt{2} & 0 & 0 \\
-\sqrt{2} & 0 & -1 & 0 \\
0 & -1 & 0 & -\sqrt{2} \\
0 & 0 & -\sqrt{2} & i\gamma /J%
\end{array}%
\right) .
\end{eqnarray}%
Our goal is to present the connections between them. To this end, the
relevant eigenvectors and eigenvalues of $\mathcal{H}$ (as well as $\mathcal{%
H}^{\dag }$) can be exactly obtained as%
\begin{equation}
\varphi _{1,2}=\left(
\begin{array}{c}
\left( a_{\pm }^{\ast }\right) ^{3} \\
\sqrt{2}a_{\pm }^{\ast } \\
\sqrt{2}a_{\pm } \\
a_{\pm }^{3}%
\end{array}%
\right) ,\varepsilon _{1,2}=\varepsilon _{\pm }=\pm J\sqrt{4-\gamma ^{2}}
\end{equation}%
and%
\begin{eqnarray}
\varphi _{3,4} &=&\left(
\begin{array}{c}
-1 \\
\frac{\sqrt{2}}{2}\left( i\gamma /J-1\right) \\
\frac{\sqrt{2}}{2}\left( i\gamma /J+1\right) \\
1%
\end{array}%
\right) ,\left(
\begin{array}{c}
1 \\
-\frac{\sqrt{2}}{2}\left( i\gamma /J+1\right) \\
\frac{\sqrt{2}}{2}\left( i\gamma /J-1\right) \\
1%
\end{array}%
\right) ,  \notag \\
\varepsilon _{3} &=&-\varepsilon _{4}=-J,
\end{eqnarray}%
where the complex numbers are
\begin{equation}
a_{\pm }=\sqrt{2}(i\gamma /J-\varepsilon _{\pm })^{-1/2}.
\end{equation}%
Here the eigenvectors are written as $\mathcal{PT}$-symmetric form, i.e., $%
\mathcal{PT}\left\vert \varphi _{i}\right\rangle =\left\vert \varphi
_{i}\right\rangle $.

And the eigenvectors and eigenvalues of $H$ can also be exactly obtained as
\begin{eqnarray}
\psi _{\lambda +} &=&\frac{1}{2}\left(
\begin{array}{c}
-e_{\lambda \pm }/J\mp 1 \\
\sqrt{2} \\
\pm \sqrt{2} \\
\mp e_{\lambda \pm }/J-1%
\end{array}%
\right) ,\lambda =1,2,  \notag \\
e_{\lambda \pm } &=&\frac{1}{2}\{[V\pm \left( \kappa -J\right) ]  \notag \\
&&+\left( -1\right) ^{\lambda }\sqrt{\left( V-\kappa -J\right) ^{2}+8J^{2}}%
\},
\end{eqnarray}%
Here the states $\left\{ \varphi _{i},i=1,2,3,4\right\} $ are also
eigenstates of the $\mathcal{PT}$\ operator. Now we apply our conclusion,
i.e., Eq\textbf{.} (\ref{Eq_A}), on states $\left\{ \varphi _{i}\right\} $
one by one and demonstrate the main point.

(i) $\varphi _{1}:$ For this state, Eq. (\ref{Eq_A}) tells us
\begin{equation}
V+\kappa =\frac{\gamma ^{2}(\sqrt{4-\gamma ^{2}}+1)}{(\sqrt{4-\gamma ^{2}}%
+2)(\sqrt{4-\gamma ^{2}}-1)},
\end{equation}%
which yields
\begin{equation}
\psi _{1+}=\frac{1}{2}\left(
\begin{array}{c}
-e_{1+}-1 \\
\sqrt{2} \\
\sqrt{2} \\
-e_{1+}-1%
\end{array}%
\right) ,e_{1+}=-\sqrt{4-\gamma ^{2}},
\end{equation}%
for $\gamma ^{2}\leqslant 3$. Then we have the relation
\begin{equation}
\varphi _{1}+\phi _{1}=2(\sqrt{4-\gamma ^{2}}+2)^{1/2}\psi _{1+}.
\end{equation}%
On the other hand, if $\gamma ^{2}\geqslant 3$, we get
\begin{equation}
\psi _{2+}=\frac{1}{2}\left(
\begin{array}{c}
-e_{2+}-1 \\
\sqrt{2} \\
\sqrt{2} \\
-e_{2+}-1%
\end{array}%
\right) ,e_{2+}=-\sqrt{4-\gamma ^{2}},
\end{equation}%
which results in
\begin{equation}
\varphi _{1}+\phi _{1}=2(\sqrt{4-\gamma ^{2}}+2)^{1/2}\psi _{2+}.
\end{equation}

(ii) $\varphi _{2}:$ For this state, by the similar procedure, we obtain
\begin{equation}
\varphi _{2}+\phi _{2}=2(2-\sqrt{4-\gamma ^{2}})^{1/2}\psi _{2+},
\end{equation}%
for $\gamma ^{2}\leqslant 3$ and
\begin{equation}
\varphi _{2}+\phi _{2}=2(2-\sqrt{4-\gamma ^{2}})^{1/2}\psi _{1+},
\end{equation}%
for $\gamma ^{2}\geqslant 3$, respectively.

(iii) $\varphi _{3}:$ For this state, Eq. (\ref{Eq_AB}) tells us $V=\kappa $%
, which leads to
\begin{equation}
\psi _{1-}=\frac{1}{\sqrt{2}}\left(
\begin{array}{c}
\sqrt{2} \\
1 \\
-1 \\
-\sqrt{2}%
\end{array}%
\right) ,e_{1-}=-1,
\end{equation}%
and%
\begin{equation}
\varphi _{3}+\phi _{3}=-2\psi _{1-}.
\end{equation}

(iv) $\varphi _{4}:$ For this state, Eq. (\ref{Eq_A}) tells us $V=-\kappa $,
which leads to
\begin{equation}
\psi _{2+}=\frac{1}{\sqrt{2}}\left(
\begin{array}{c}
-\sqrt{2} \\
1 \\
1 \\
-\sqrt{2}%
\end{array}%
\right) ,e_{2+}=1,
\end{equation}%
and correspondingly
\begin{equation}
\varphi _{4}+\phi _{4}=-2\psi _{2+}.
\end{equation}

Based on the explicit solutions, we conclude that for a given $\gamma $, $%
\varphi _{i}+\phi _{i}$, a superposition of real-energy eigenstates of $%
\mathcal{H}$\ and $\mathcal{H}^{\dag }$, always corresponds to an eigenstate
$\psi _{j}$\ of $H$, i.e.,%
\begin{equation}
\varphi _{i}+\phi _{i}\propto \psi _{j},
\end{equation}%
with the same eigenenergy. These facts demonstrate and verify our analysis
in the last section. Moreover, it also has an implication that one can find
the corresponding Hermitian Hamiltonian for every state $\varphi _{i}+\phi
_{i}$.

For large $N$, it is a little difficult to give analytical expressions of
eigenstates for arbitrary parameters. Fortunately, we can provide some
eigenstates for specific parameters. We consider the case with $N=4m+3$ ($m$
is an integer), $\gamma =\pm 2J$, and $\kappa =V$. For non-Hermitian
Hamiltonians $\mathcal{H}$\ and $\mathcal{H}^{\dag }$, there is a
zero-energy state, which is the coalescing state of three levels with
eigenenergies $0$, and $\pm \sqrt{4J^{2}-\gamma ^{2}}$ for $\gamma
^{2}\leqslant 4J^{2}$, respectively. The eigenstates of $\mathcal{H}$\ and $%
\mathcal{H}^{\dag }$ can be written as%
\begin{eqnarray}
\left\vert \Phi _{-}\right\rangle &=&\left\vert \Phi _{+}\right\rangle
^{\ast }=\frac{1}{\sqrt{2\left( N-1\right) }}[\left\vert A\right\rangle
\notag \\
&&-i^{N+1}\left\vert B\right\rangle -\sqrt{2}\sum_{l=2}^{N-1}i^{l+1}\left%
\vert l\right\rangle ],
\end{eqnarray}%
which can be checked to satisfy $\mathcal{H}\left\vert \Phi
_{-}\right\rangle =\mathcal{H}^{\dag }\left\vert \Phi _{+}\right\rangle =0$
and $\langle \Phi _{\pm }\left\vert \Phi _{\pm }\right\rangle =1$. We note
that the biorthogonal norm of $\left\vert \Phi _{\pm }\right\rangle $\ is%
\begin{equation}
\langle \Phi _{+}\left\vert \Phi _{-}\right\rangle =0,
\end{equation}%
which indicates that it is a coalescing state. On the other hand, the
zero-energy state of $H$\ is%
\begin{equation}
\left\vert \Psi \right\rangle =\frac{1}{\sqrt{N-1}}(\sum_{j=2}^{N-1}\sin
\frac{\pi j}{2}\left\vert j\right\rangle +\sqrt{2}\left\vert 1\right\rangle -%
\sqrt{2}\left\vert N\right\rangle ),
\end{equation}%
which satisfies $H\left\vert \Psi \right\rangle =0$. And it is easy to check
that%
\begin{equation}
\left\vert \Psi \right\rangle =\left\vert \Phi _{+}\right\rangle +\left\vert
\Phi _{-}\right\rangle .
\end{equation}%
The physical picture of this relation is clear: states $\left\vert \Phi
_{+}\right\rangle $ and $\left\vert \Phi _{-}\right\rangle $ represent two
plane waves with wave vectors $\pm \pi /2$, while\ state $\left\vert \Psi
\right\rangle $\ stands for a standing wave.

\subsection{Su-Schrieffer-Heeger chain}

\label{Long-range entanglement}

\begin{figure}[htb]
\centering
\includegraphics[ bb=49 516 486 758, width=0.4\textwidth, clip]{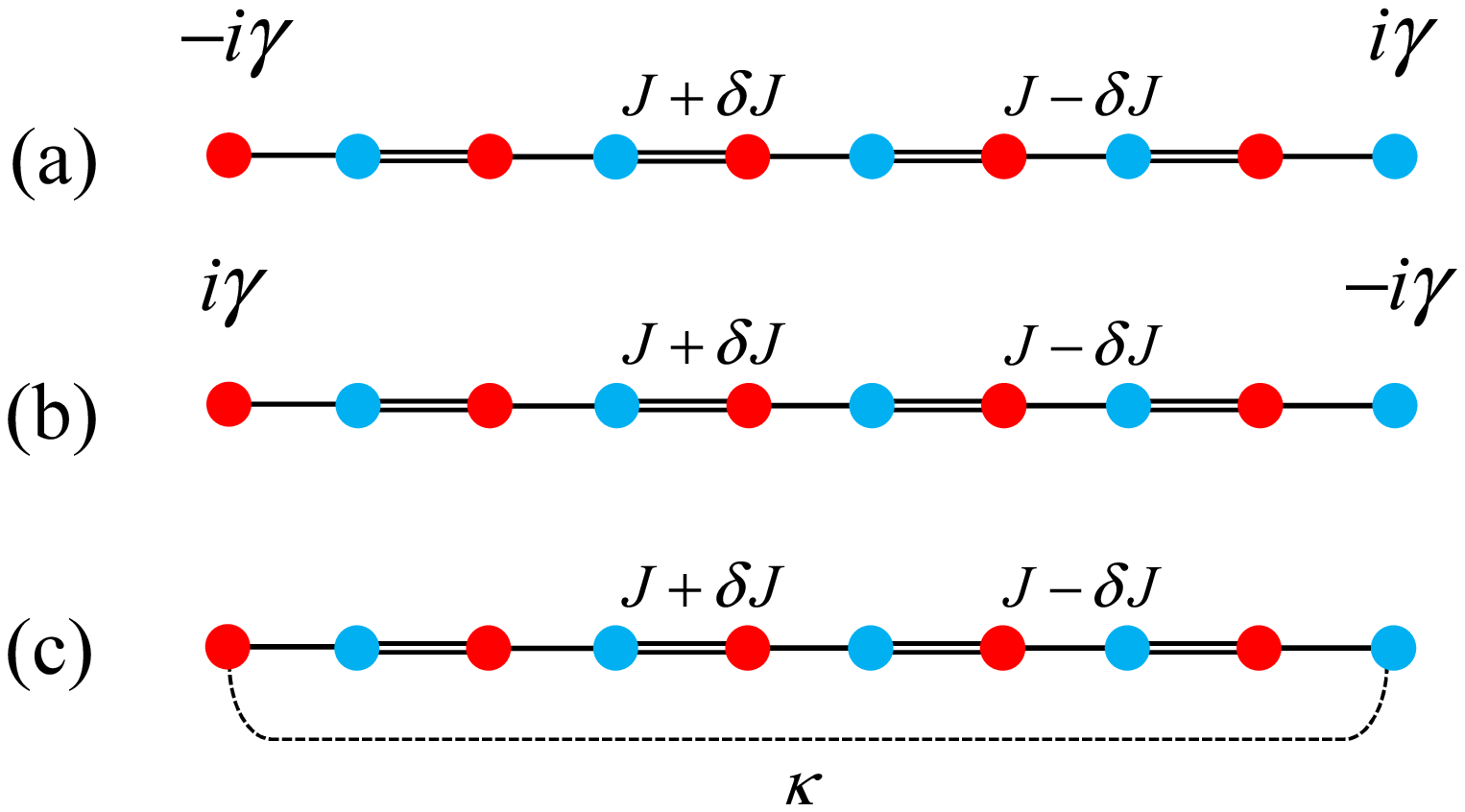}
\caption{ Schematic illustrations for the SSH chain. Here $H_{\text{\textrm{%
sub}}}$ is an SSH chain of length $N$. The combination of zero-mode
eigenstates of systems (a) and (b) is also zero-mode eigenstate of system
(c) under the condition $\protect\kappa =\protect\gamma =J\left( 1+\protect%
\delta \right) \Delta ^{N/2},\Omega =\protect\sqrt{2\protect\delta %
J^{2}/(J^{2}(1+\protect\delta )^{2}-\protect\kappa ^{2})}$. }
\label{fig3}
\end{figure}

In this section, we present another example to show the application of above
conclusion for quantum engineering. We take $H_{\text{sub}}$\ to be an SSH
chain, which is proposed by Su, Schrieffer, and Heeger (SSH) to model
polyacetylene \cite{Su,Schrieffer}, is the prototype of a topologically
nontrivial band insulator with a symmetry protected topological phase \cite%
{Ryu,Wen}. In recent years, it has been attracted much attention and
extensive studies have been demonstrated \cite%
{Xiao,Hasan,X.L.Q,Delplace,ChenS1,ChenS2}. The sub-Hamiltonian in this
example is%
\begin{eqnarray}
H_{\text{sub}} &=&-\left( J-J\delta \right) \sum_{j=1}^{N/2}\left\vert
2j-1\right\rangle \left\langle 2j\right\vert \\
&&-\left( J+J\delta \right) \sum_{j=1}^{N/2-j}\left\vert 2j\right\rangle
\left\langle 2j+1\right\vert +\mathrm{H.c.},
\end{eqnarray}%
and non-Hermitian term is%
\begin{equation}
H_{\gamma }=-i\gamma \left( \left\vert 1\right\rangle \left\langle
1\right\vert -\left\vert N\right\rangle \left\langle N\right\vert \right) ,
\end{equation}%
and Hermitian term is%
\begin{equation}
H_{\kappa }=-\kappa (\left\vert 1\right\rangle \left\langle N\right\vert +%
\mathrm{H.c.}),
\end{equation}%
which are sketched in Fig. \ref{fig3}. We note that $\mathcal{H}$($\mathcal{H%
}^{\dag }$)\ also has $\mathcal{PT}$ symmetry. It is tough to give an
explicit form of the eigenstates. Fortunately, exact zero-mode eigenstates
of both $H$ and $\mathcal{H}$($\mathcal{H}^{\dag }$)\ are obtained for
specific relation among $\gamma $, $\delta $\ and $\kappa $\ \cite{Lin}. For
Hamiltonian $H$, there are two degenerate zero-mode eigenstates, which has
the form%
\begin{equation}
\left\vert \psi _{\pm }\right\rangle =\Omega \sum_{j=1}^{N/2}[\left( -\Delta
\right) ^{j-1}\left\vert 2j-1\right\rangle \pm \left( -\Delta \right)
^{N/2-j}\left\vert 2j\right\rangle ],
\end{equation}%
\begin{eqnarray}
\left\vert \psi _{1}\right\rangle &=&\sqrt{2}\Omega \sum_{j=1}^{N/2}\left(
-\Delta \right) ^{j-1}\left\vert 2j-1\right\rangle , \\
\left\vert \psi _{2}\right\rangle &=&\sqrt{2}\Omega \sum_{j=1}^{N/2}\left(
-\Delta \right) ^{N/2-j}\left\vert 2j\right\rangle ,
\end{eqnarray}%
satisfying
\begin{equation}
H\left( \kappa _{\mathrm{c}}\right) \left\vert \psi _{1}\right\rangle
=[H\left( \kappa _{\mathrm{c}}\right) ]^{\dagger }\left\vert \psi
_{2}\right\rangle =0.
\end{equation}%
and%
\begin{equation}
\kappa =\kappa _{\mathrm{c}}=J(1+\delta )\Delta ^{N/2},
\end{equation}%
where $\Delta =\left( 1-\delta \right) /\left( 1+\delta \right) $\ denotes
the staggered hopping strength and $\Omega =\sqrt{2\delta
J^{2}/(J^{2}(1+\delta )^{2}-\kappa ^{2})}$ is the Dirac normalizing
constant. For non-Hermitian Hamiltonian $\mathcal{H}$($\mathcal{H}^{\dag }$%
), the zero-mode state is a coalescence state

\begin{equation}
\left\vert \varphi _{\mathrm{zm}}\right\rangle =\Omega
\sum_{j=1}^{N/2}[\left( -\Delta \right) ^{j-1}\left\vert 2j-1\right\rangle
+i\left( -\Delta \right) ^{N/2-j}\left\vert 2j\right\rangle ]\left\vert
\text{\textrm{vac}}\right\rangle ,
\end{equation}%
Similarly, the zero-mode state for $H_{\gamma _{\mathrm{c}}}^{\dagger }$ can
be constructed as%
\begin{equation}
\left\vert \eta _{\mathrm{zm}}\right\rangle =\Omega \sum_{j=1}^{N/2}[\left(
-\Delta \right) ^{j-1}\left\vert 2j-1\right\rangle -i\left( -\Delta \right)
^{N/2-j}\left\vert 2j\right\rangle ]\left\vert \text{\textrm{vac}}%
\right\rangle ,
\end{equation}%
satisfying%
\begin{equation}
H\left( \gamma _{\mathrm{c}}\right) \left\vert \varphi _{\mathrm{zm}%
}\right\rangle =[H\left( \gamma _{\mathrm{c}}\right) ]^{\dagger }\left\vert
\eta _{\mathrm{zm}}\right\rangle =0.
\end{equation}%
and%
\begin{equation}
\gamma =\gamma _{\mathrm{c}}=\kappa _{\mathrm{c}}=J(1+\delta )\Delta ^{N/2},
\label{Gamma_Condition}
\end{equation}

\begin{figure}[tbh]
\includegraphics[ bb=87 168 541 602, width=0.23\textwidth, clip]{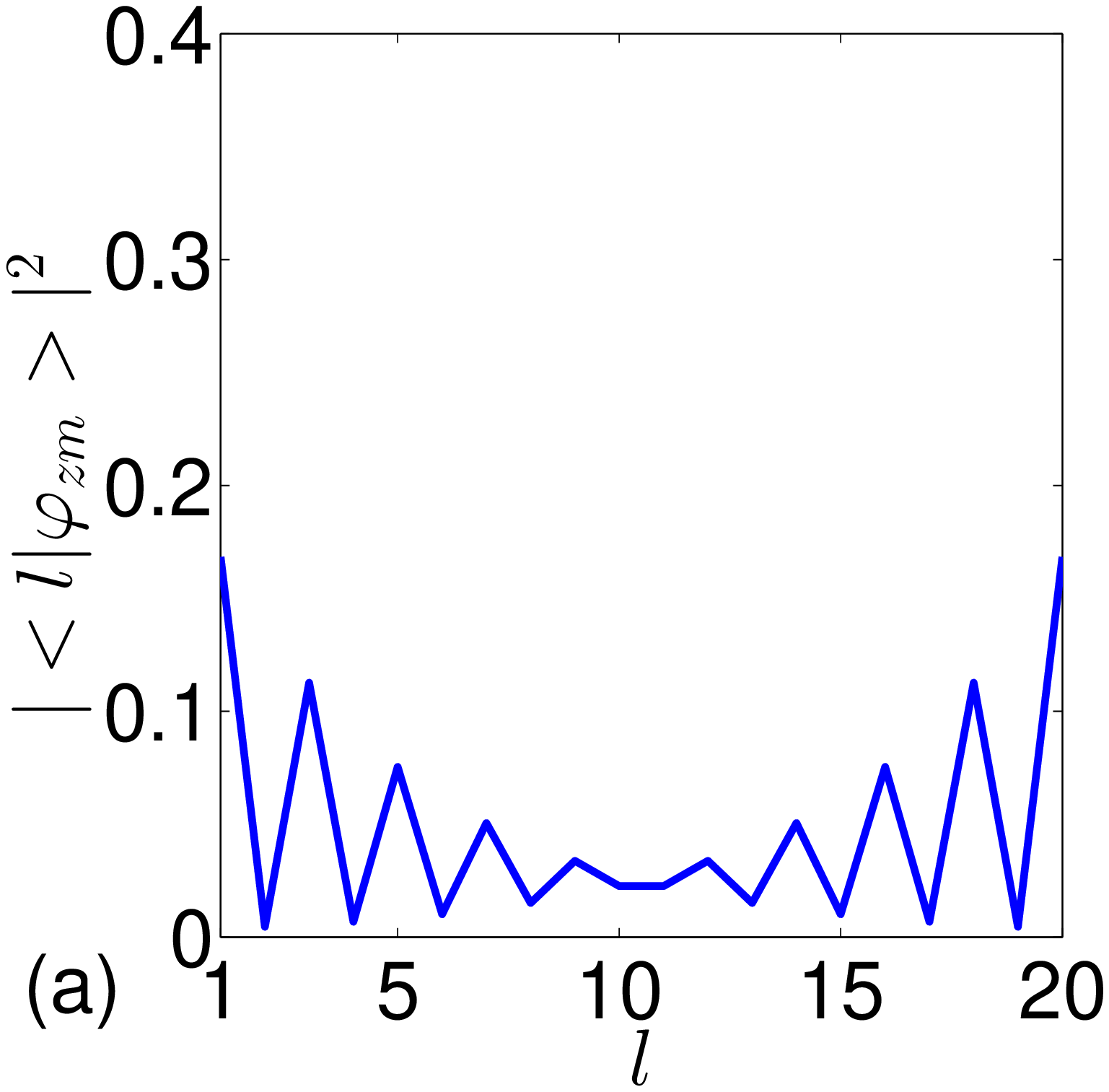} %
\includegraphics[ bb=87 168 541 602, width=0.23\textwidth, clip]{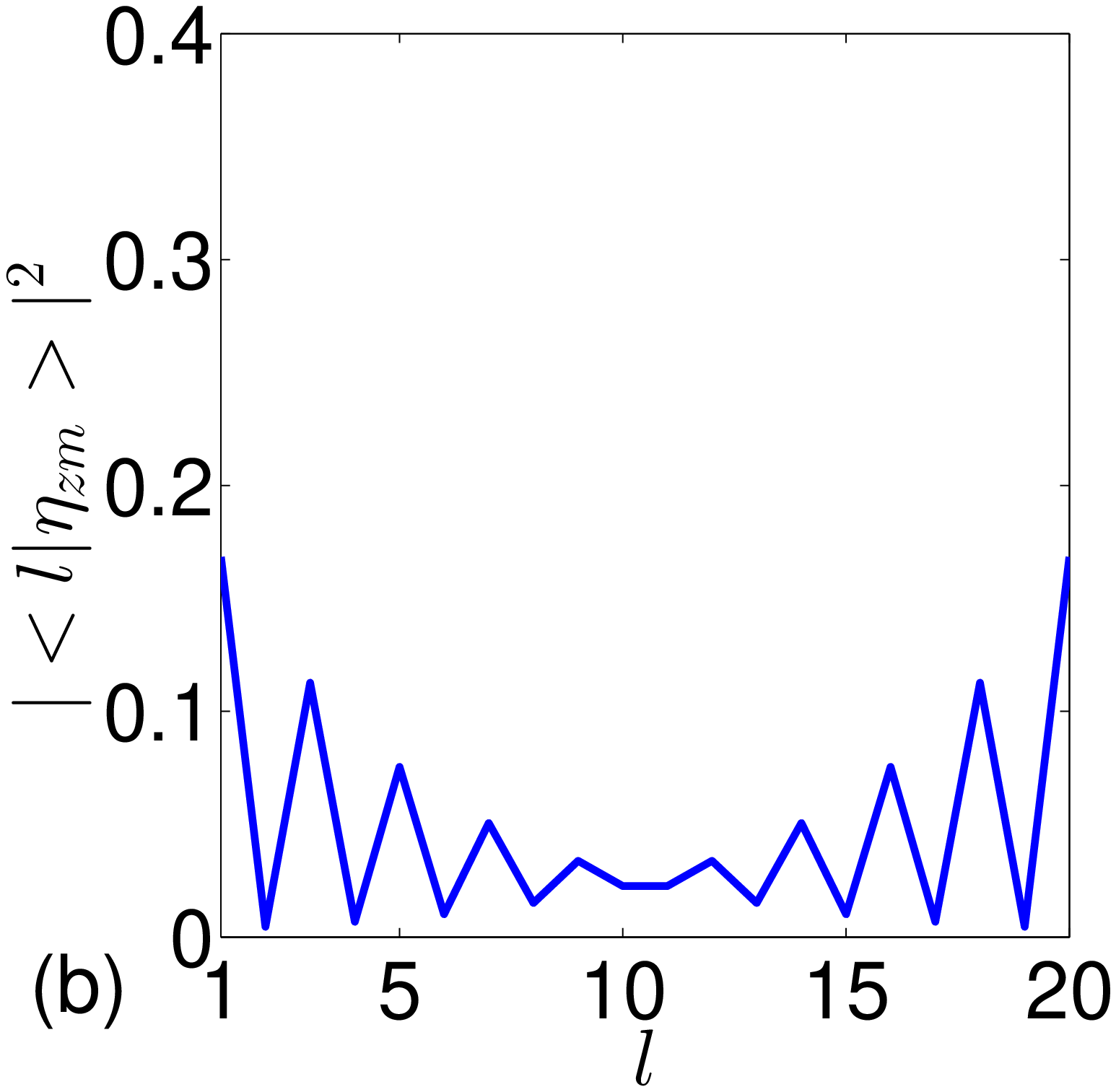} %
\includegraphics[ bb=87 168 541 602, width=0.23\textwidth, clip]{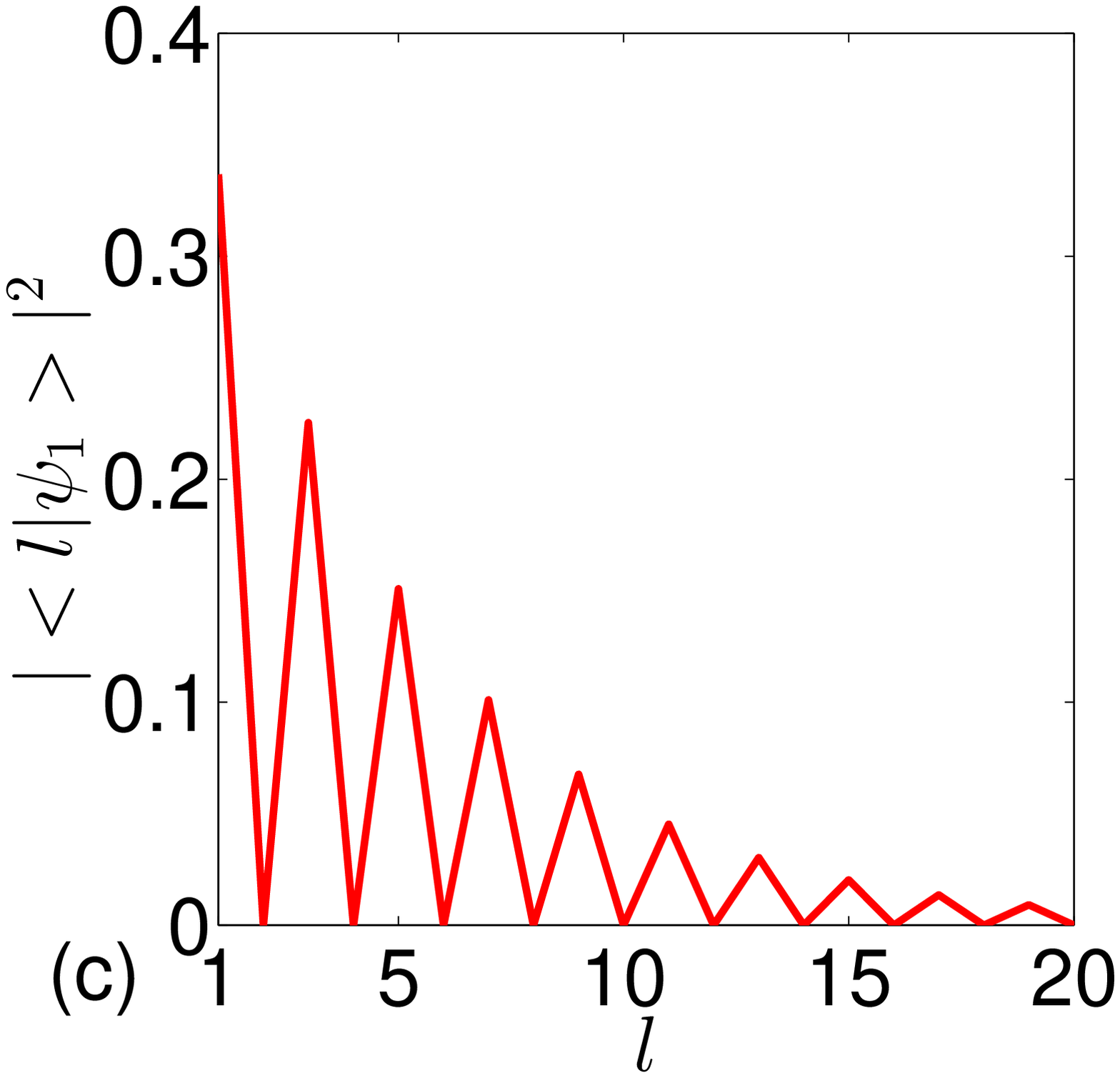} %
\includegraphics[ bb=87 168 541 602, width=0.23\textwidth, clip]{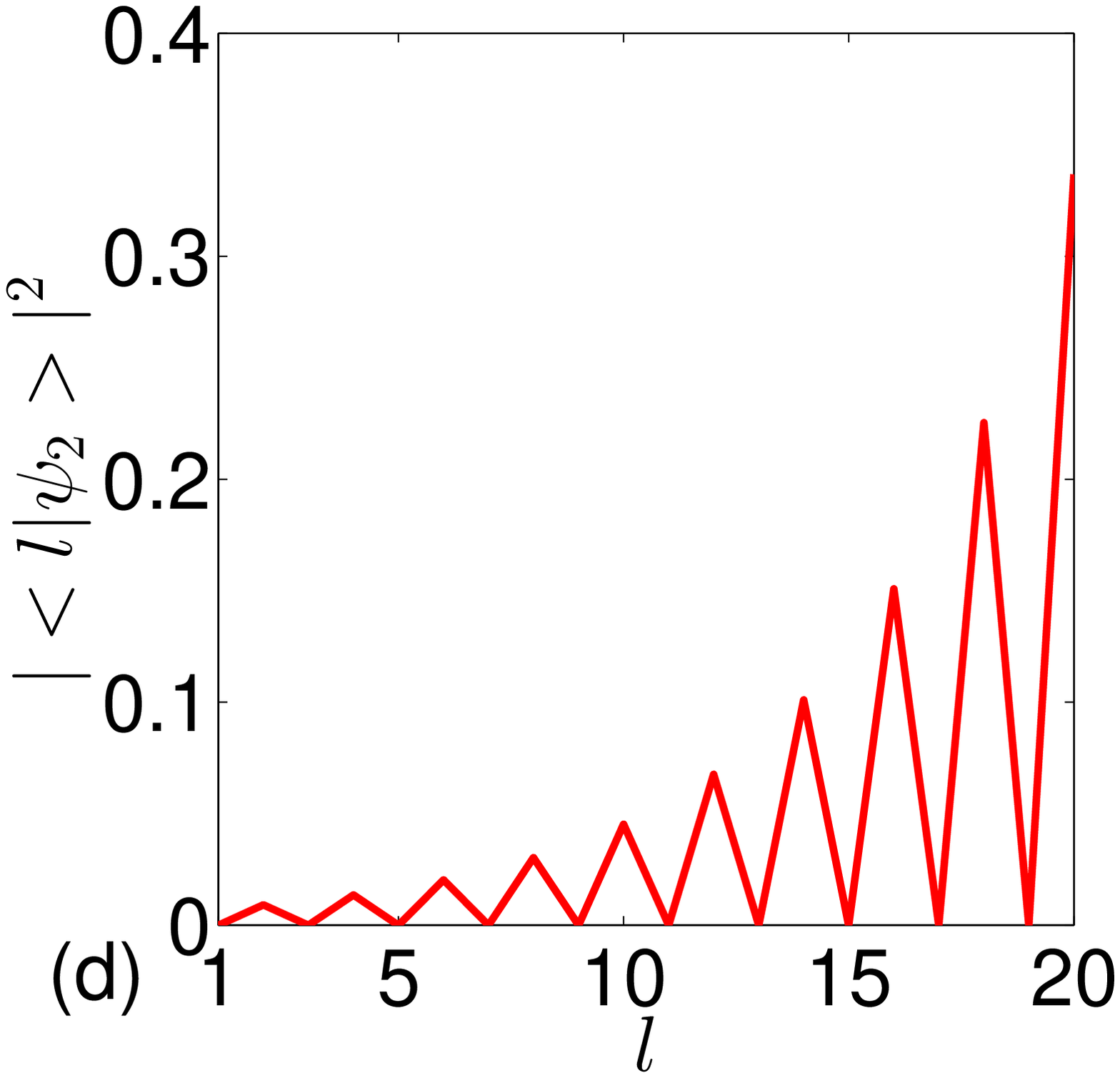}
\caption{ The Dirac probabilities of the four\ zero-mode eigstates for SSH
model in Fig. 4(a-c) at $\protect\delta =0.1,N=20$. (a) $\langle \protect%
\varphi _{\mathrm{zm}}\left\vert \protect\varphi _{\mathrm{zm}}\right\rangle
$, (b) $\langle \protect\eta _{\mathrm{zm}}\left\vert \protect\eta _{\mathrm{%
zm}}\right\rangle $, (c) $\langle \protect\psi _{1}\left\vert \protect\psi %
_{1}\right\rangle $, (d) $\langle \protect\psi _{2}\left\vert \protect\psi %
_{2}\right\rangle $. }
\label{fig4}
\end{figure}
Three wave functions for finite $N=20$ are plotted in Fig. \ref{fig4}. They
satisfies the following relations%
\begin{eqnarray*}
\left\vert \varphi _{\mathrm{zm}}\right\rangle +\left\vert \eta _{\mathrm{zm}%
}\right\rangle  &=&\sqrt{2}\left\vert \psi _{1}\right\rangle , \\
\left\vert \varphi _{\mathrm{zm}}\right\rangle -\left\vert \eta _{\mathrm{zm}%
}\right\rangle  &=&\sqrt{2}\left\vert \psi _{2}\right\rangle .
\end{eqnarray*}%
The first relationship is accords with our previous conclusion, which shows
a symmetric combination under relation $\left\vert \varphi _{\mathrm{zm}%
}\right\rangle +\left\vert \eta _{\mathrm{zm}}\right\rangle $; the second
relationship shows an anti-symmetric combination of the eigen states under
relation $\left\vert \varphi _{\mathrm{zm}}\right\rangle -\left\vert \eta _{%
\mathrm{zm}}\right\rangle $. The results of two concrete examples show that
our general conclusion is feasible in practice.
\begin{figure*}[tbph]
\centering
\includegraphics[ bb=82 188 492 583, width=0.3\textwidth, clip]{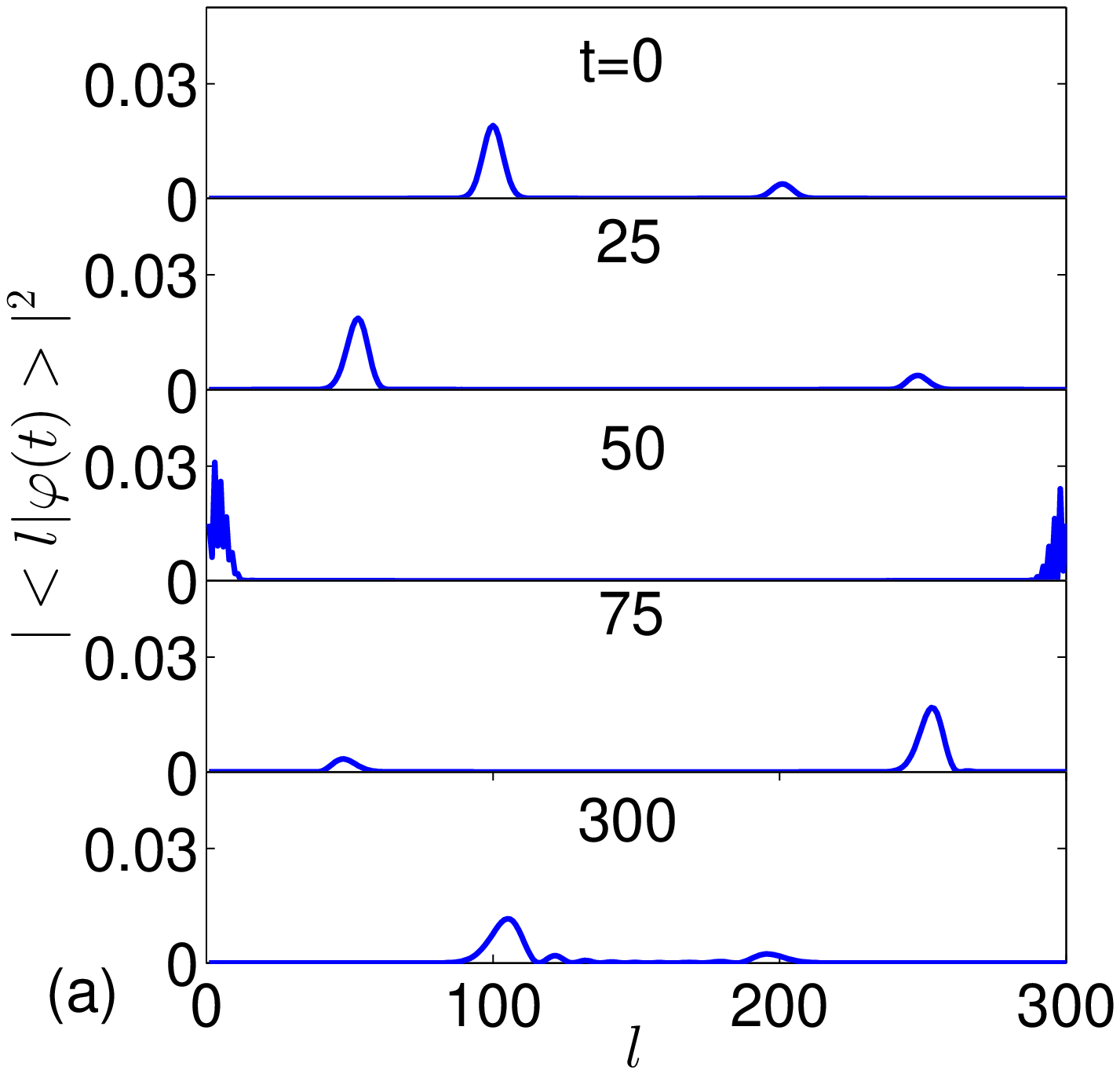} %
\includegraphics[ bb=82 188 492 583, width=0.3\textwidth, clip]{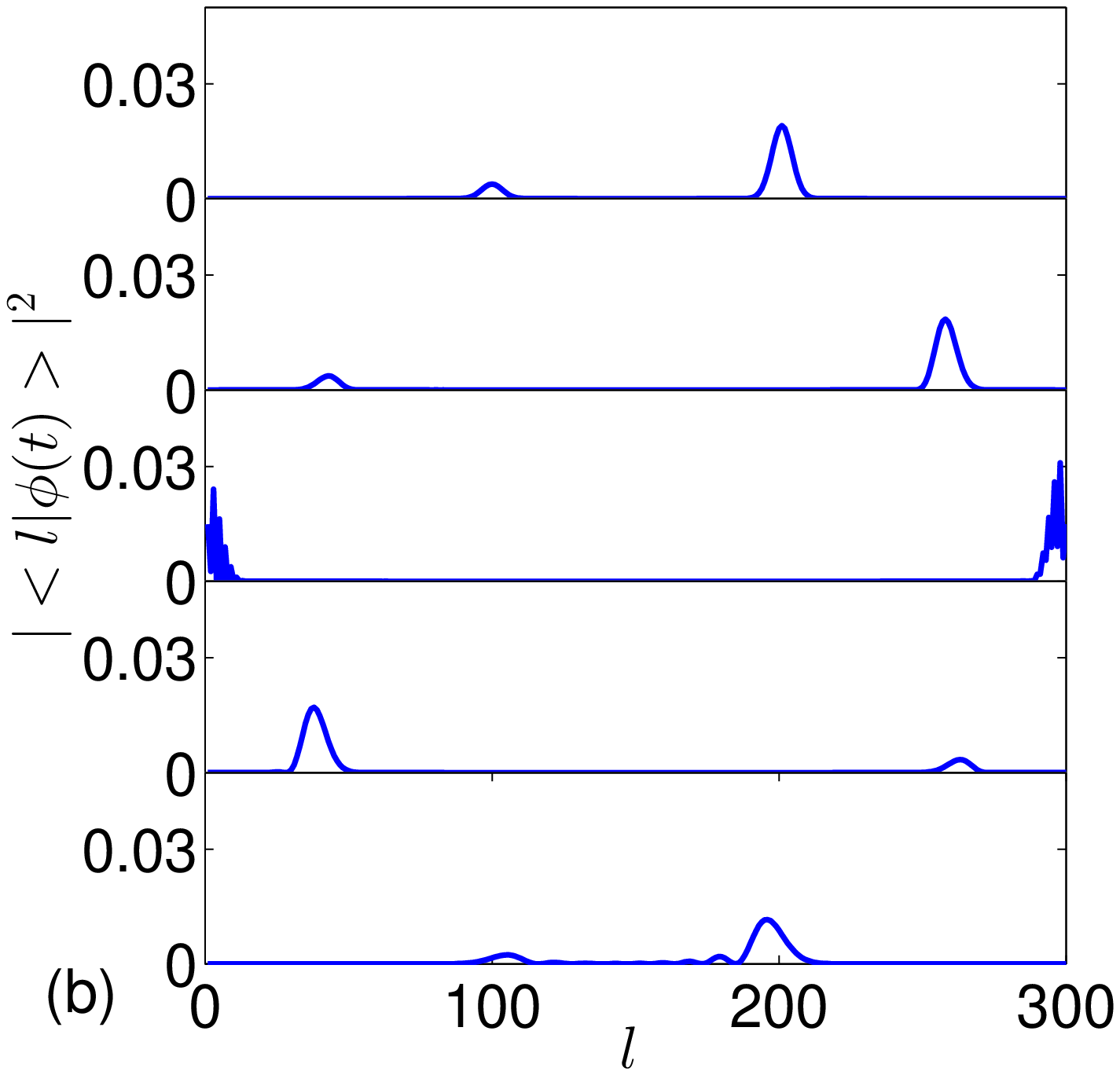} %
\includegraphics[ bb=82 188 492 583, width=0.3\textwidth, clip]{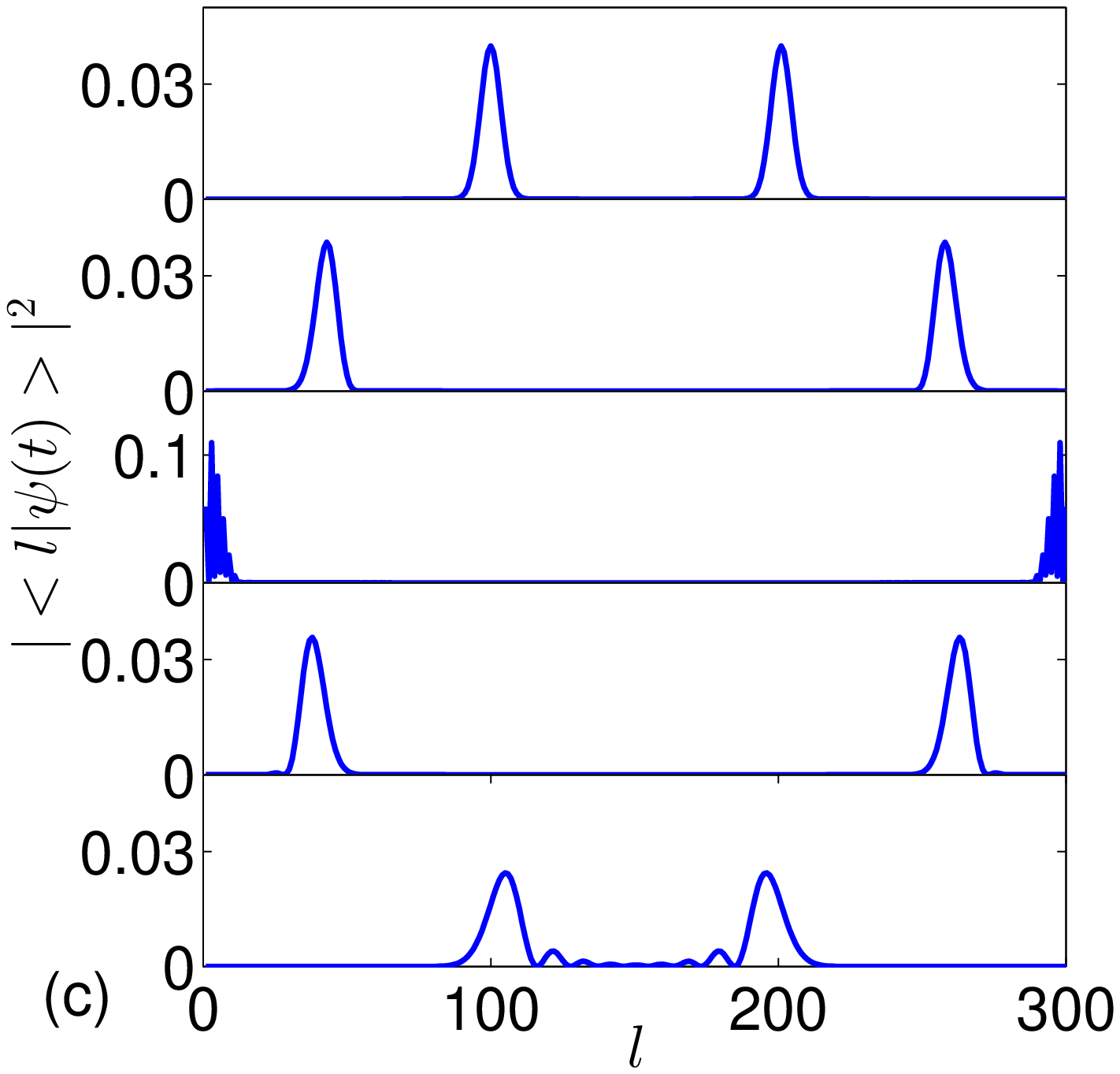}
\caption{ The time evolution profiles of three initial states with the same
distribution in Eq. (\protect\ref{initial state}) with $\protect\alpha =0.2$%
, the evolutions are governed by their own systems. (a) The initial state
state $\left\vert \protect\varphi (0)\right\rangle $ is in $\mathcal{H}$,
(b) The initial state $\left\vert \protect\phi (0)\right\rangle $ is in $%
\mathcal{H}^{\dag }$, (c) The initial state $\left\vert \protect\psi \left(
0\right) \right\rangle $ or $\left\vert \protect\varphi (0)\right\rangle
+\left\vert \protect\phi (0)\right\rangle $ in $H$. The unit of time is $1/J$%
. The plots are taken for five typical instants. In (c), \textbf{$\left\vert
\left\vert \protect\psi \left( t\right) \right\rangle \right\vert ^{2}$ }is
plotted in order to compare with that in (a) and (b). }
\label{fig5}
\end{figure*}

\section{Parallel dynamics}

\label{Parallel dynamics}

In this section, we apply the obtained\ results on the dynamics of the
relevant Hamiltonians. In the above section, we have shown that there exist
three sets of eigenstates $\left\{ \left\vert \psi _{n}\right\rangle
\right\} $, $\left\{ \left\vert \varphi _{n}\right\rangle \right\} $, and $%
\left\{ \left\vert \phi _{n}\right\rangle \right\} $ for the Hamiltonians $H$%
, $\mathcal{H}$, and $\mathcal{H}^{\dag }$, respectively, which obey the
relation $\left\vert \psi _{n}\right\rangle =\left\vert \varphi
_{n}\right\rangle +\left\vert \phi _{n}\right\rangle $. The three sets of
eigenstates have the same eigenenergies $\left\{ \varepsilon _{n}\right\} $.
Now we consider the time evolution of an initial state in the subspace $%
\left\{ \left\vert \psi _{n}\right\rangle \right\} $, i.e.,%
\begin{equation}
\left\vert \psi (0)\right\rangle =\sum_{n}c_{n}\left\vert \psi
_{n}\right\rangle .
\end{equation}%
The evolved state is%
\begin{eqnarray}
\left\vert \psi (t)\right\rangle &=&\sum_{n}c_{n}e^{-i\varepsilon
_{n}t}\left\vert \psi _{n}\right\rangle  \notag \\
&=&e^{-i\mathcal{H}t}\left\vert \varphi (0)\right\rangle +e^{-i\mathcal{H}%
^{\dag }t}\left\vert \phi (0)\right\rangle ,
\end{eqnarray}%
where
\begin{equation}
\left\vert \varphi (0)\right\rangle =\sum_{n}c_{n}\left\vert \varphi
_{n}\right\rangle ,\left\vert \phi (0)\right\rangle =\sum_{n}c_{n}\left\vert
\phi _{n}\right\rangle .
\end{equation}%
\ It indicates that, for a dynamic process $\left\vert \varphi
(t)\right\rangle $\ in a non-Hermitian system $\mathcal{H}$, there always
exists a parallel dynamic process\ $\left\vert \phi (t)\right\rangle $\
governed by the corresponding Hermitian conjugate Hamiltonian $\mathcal{H}%
^{\dag }$. And it is shown that a linear superposition of the two parallel
dynamics may be exactly equivalent to the time evolution of a state $%
\left\vert \psi (t)\right\rangle $\ under a Hermitian Hamiltonian $H$.

Furthermore, we note that, the total Dirac probabilities of three states
obey the relations
\begin{equation}
\langle \psi (t)\left\vert \psi (t)\right\rangle =\langle \varphi
(t)\left\vert \varphi (t)\right\rangle +\langle \phi (t)\left\vert \phi
(t)\right\rangle +2\theta ,
\end{equation}%
and
\begin{equation}
\frac{\text{d}}{\text{d}t}\left\vert \left\vert \varphi (t)\right\rangle
\right\vert ^{2}+\frac{\text{d}}{\text{d}t}\left\vert \left\vert \phi
(t)\right\rangle \right\vert ^{2}=\frac{\text{d}}{\text{d}t}\left\vert
\left\vert \psi (t)\right\rangle \right\vert ^{2}=0,
\end{equation}%
due to the biorthonormal relation%
\begin{equation}
\frac{1}{\theta }\langle \varphi (t)\left\vert \phi (t)\right\rangle =\frac{1%
}{\theta }\langle \varphi (t)\left\vert \phi (t)\right\rangle =1,
\end{equation}%
where $\theta $ is a constant.

Specifically, when the system is $\mathcal{PT}$-symmetric and $\mathcal{H}%
^{\dag }=\mathcal{H}^{\ast }$, we have $\mathcal{PHP}^{-1}=\mathcal{H}^{\dag
}$. We note that operation $\mathcal{P}$\ cannot affect the Dirac
probability, i.e., $\left\vert \left\vert \varphi (t)\right\rangle
\right\vert ^{2}=\left\vert \left\vert \phi (t)\right\rangle \right\vert
^{2} $. Then we have%
\begin{equation}
\frac{\text{d}}{\text{d}t}\left\vert \left\vert \varphi (t)\right\rangle
\right\vert ^{2}=\frac{\text{d}}{\text{d}t}\left\vert \left\vert \phi
(t)\right\rangle \right\vert ^{2}=0,
\end{equation}%
i.e., the Dirac probabilities of the two states are conservative. In
general, a non-Hermitian system obeys the conservation of biorthogonal
probability rather than Dirac probability. However, in the sub-sets $\left\{
\left\vert \varphi _{n}\right\rangle \right\} $ and $\left\{ \left\vert \phi
_{n}\right\rangle \right\} $, both types of probability are conservative.

Next we demonstrate this feature by the numerical simulation of the time
evolution of a specific initial state under a concrete system. We consider
the Hamiltonians $H$, $\mathcal{H}$, and $\mathcal{H}^{\dag }$, which
consist of sub-Hamiltonians $H_{\text{sub}}$, $H_{\gamma }$, and $H_{\kappa
} $ defined in Eqs. (\ref{H_sub}), (\ref{H_gamma}), and (\ref{H_kappa}),
respectively. Here the parameters are taken as $N=300$, $\gamma =0.75$, and $%
\kappa =-V=-1$. In this case, an exact eigenstate can be obtained as
\begin{eqnarray}
\left\vert \psi _{0}\right\rangle &=&\frac{1}{\sqrt{2\left( N-1\right) }}%
\left( \overset{N-1}{\underset{l=2}{\sum }}\sqrt{2}\left\vert l\right\rangle
+\left\vert 1\right\rangle +\left\vert N\right\rangle \right) \\
\varepsilon _{0} &=&-2J.
\end{eqnarray}%
On the other hand, numerical result shows that there are three $149$%
-dimensional sets of eigenstates $\{H:\left\vert \psi _{m}\right\rangle \}$,
$\{\mathcal{H}:\left\vert \varphi _{m}\right\rangle \}$, and $\{\mathcal{H}%
^{\dag }:\left\vert \phi _{m}\right\rangle \}$, which have the identical
eigenenergies $\left\{ \varepsilon _{m}\right\} $, $m\in \lbrack 1,149]$.
State $\left\vert \psi _{0}\right\rangle $\ is not included in the set $%
\{\left\vert \psi _{m}\right\rangle \}$. These states satisfy%
\begin{eqnarray}
\mathcal{P}\left\vert \psi _{m}\right\rangle &=&\left\vert \psi
_{m}\right\rangle , \\
\mathcal{PT}\left\vert \varphi _{m}\right\rangle &=&\left\vert \varphi
_{m}\right\rangle , \\
\mathcal{PT}\left\vert \phi _{m}\right\rangle &=&\left\vert \phi
_{m}\right\rangle ,
\end{eqnarray}%
and can be written as the form
\begin{eqnarray}
\langle A\left\vert \varphi _{m}\right\rangle &=&\langle A\left\vert \phi
_{m}\right\rangle =\langle A\left\vert \psi _{m}\right\rangle /2 \\
&=&\langle B\left\vert \varphi _{m}\right\rangle =\langle B\left\vert \phi
_{m}\right\rangle =\langle B\left\vert \psi _{m}\right\rangle /2.
\end{eqnarray}%
Then we have the relation $\left\vert \psi _{m}\right\rangle =\left\vert
\varphi _{m}\right\rangle +\left\vert \phi _{m}\right\rangle $. Based on
these analyses, we find that an arbitrary state $\left\vert \psi \left(
0\right) \right\rangle $\ satisfying%
\begin{equation}
\mathcal{P}\left\vert \psi \left( 0\right) \right\rangle =\left\vert \psi
\left( 0\right) \right\rangle ,\langle \psi _{0}\left\vert \psi \left(
0\right) \right\rangle =0,
\end{equation}%
can be the state exhibiting the parallel dynamics. In this paper, we
consider an Gaussian wave packet
\begin{equation}
\left\vert N/3,\pi /2\right\rangle =\frac{1}{\sqrt{\Omega }}%
\sum_{j}e^{-\alpha ^{2}\left( j-N/3\right) ^{2}}e^{i\pi /2j}\left\vert
j\right\rangle .
\end{equation}%
Here $\Omega =\sqrt{\pi /2}/\alpha $\ is the normalization factor and $\sqrt{%
2\ln 2}/\alpha \ll N/2$\ is the width\textbf{\ }of the wave packet. We take
the central momentum of the wave packet as $\pi /2$, which weeds out\textbf{%
\ }the component of state $\left\vert \psi _{0}\right\rangle $. The initial
state is constructed as%
\begin{equation}
\left\vert \psi \left( 0\right) \right\rangle =\frac{1}{\sqrt{2}}(\left\vert
N/3,\pi /2\right\rangle +\mathcal{P}\left\vert N/3,\pi /2\right\rangle ),
\end{equation}%
which is spanned by the set $\{\left\vert \psi _{m}\right\rangle \}$\ with
the coefficient
\begin{equation}
c_{n}=\langle \psi _{n}\left\vert \psi \left( 0\right) \right\rangle .
\label{initial state}
\end{equation}%
Based on $\{c_{n}\}$, the initial states $\left\vert \varphi
(0)\right\rangle $ and $\left\vert \phi (0)\right\rangle $ can be obtained
accordingly. We compute the time evolutions of states $\left\vert \psi
\left( 0\right) \right\rangle $, $\left\vert \varphi (0)\right\rangle $, and
$\left\vert \phi (0)\right\rangle $ by exact diagonalizations. Plots of
Dirac probabilities $\left\vert \langle l\left\vert \varphi (t)\right\rangle
\right\vert ^{2}$, $\left\vert \langle l\left\vert \phi (t)\right\rangle
\right\vert ^{2}$, and $\left\vert \langle l\left\vert \psi (t)\right\rangle
\right\vert ^{2}$\ of the evolved states at several typical instants are
listed in Fig. \ref{fig5}. The profiles of initial states are two separated
symmetric Gaussian wave packets for $H$ but asymmetric for $\mathcal{H}$ and
$\mathcal{H}^{\dag }$. At beginning, three initial states evolve in the same
way before they collide with the boundaries. It is due to the fact that
three Hamiltonians $H$, $\mathcal{H}$, and $\mathcal{H}^{\dag }$ contain the
same sub-Hamiltonian $H_{\text{sub}}$. When the wave packets reach the
boundaries, the imaginary potentials are taking effect. The positive
imaginary potentials increase the probabilities, while the negative ones
decrease the probabilities, violating the conservation of probabilities of
each individual wave packet. In contrast, the profile of $\left\vert \langle
l\left\vert \psi (t)\right\rangle \right\vert ^{2}$ or $\left\vert
\left\langle l\right\vert [\left\vert \varphi (t)\right\rangle +\left\vert
\phi (t)\right\rangle ]\right\vert ^{2}$ exhibits a Hermitian dynamical
behavior. Two symmetric wave packets collide with each other at the joint.
According to the analysis above, the total Dirac probabilities of states $%
\left\vert \varphi (t)\right\rangle $ and $\left\vert \phi (t)\right\rangle $%
\ are still conservative. This non-intuitive behavior arises from the
asymmetry of two wave packets. The gain of the small wave packet counteracts
the loss of the big one.

\section{Conclusion and discussion}

\label{Conclusion and discussion}

In conclusion, we have presented a novel way of finding the link between a
non-Hermitian Hamiltonian and a Hermitian one, based on the exact solutions.
We have found that there is a class of non-Hermitian Hamiltonians which has
a subtle relation to a class of Hermitian Hamiltonians. Unlike the previous
works, the connection refers to all the eigenenergies and eigenvectors of
three Hamiltonians. The correspondence among the three Hamiltonians is not
only for an individual state but a subset of eigenstates. The identities
about eigenvectors and eigenenergies would ensure the identical dynamics. In
this work, we just reveal the existence in tractable models (restricted in
the systems with imaginary potentials, being probably $\mathcal{PT}$ type).
We think the extension to more general models is possible. This finding
implies that there may be three parallel worlds around us: an event we
observe in our Hermitian world is the combination of two events from the
other two non-Hermitian worlds.

\section*{Acknowledgment}

We acknowledge the support of the National Natural Science Foundation of
China (Grant Nos. 11374163 and 11605094) and the Tianjin Natural Science
Foundation (Grant No. 16JCYBJC40800).

\end{document}